\documentclass[aps,prb,twocolumn,superscriptaddress,floatfix,longbibliography]{revtex4-2}

\usepackage{graphicx}
\usepackage{braket}
\usepackage{bm}
\usepackage{upgreek}
\usepackage{tabularx}
\usepackage[normalem]{ulem}
\usepackage[utf8]{inputenc}
\usepackage{mathtools}
\usepackage{array}
\usepackage{physics}

\usepackage{hyperref}

\begin{document}
\title{Bound states in the continuum in cuprous oxide quantum wells}
\author{Angelos Aslanidis}
\author{Jörg Main}
\email[Email: ]{main@itp1.uni-stuttgart.de}
\author{Patric Rommel}
\affiliation{Institut für Theoretische Physik I, Universität Stuttgart, 70550 Stuttgart, Germany}
\author{Stefan Scheel}
\author{Pavel A. Belov}
\affiliation{Institut für Physik, Universität Rostock,
Albert-Einstein-Straße 23-24, 18059 Rostock, Germany}

\date{\today}

\begin{abstract}
We propose a realistic semiconductor system containing bound states in the continuum (BICs) 
which allows for a practical realization.
By varying the confinement strength of excitons in cuprous oxide quantum wells, we show
that long-lived Rydberg states of the confined electron-hole pairs appear in the 
continuum background.
The accuracy of calculations of the linewidths based on the coupled-channel Schr\"{o}dinger 
equation with three channels and only few basis states is confirmed by a numerically exact 
solution employing a B-spline basis and the complex coordinate-rotation method.
We argue that finite-sized cuprous oxide crystals, due to their large exciton binding 
energies, are a convenient platform for experimental identification of BICs.
\end{abstract}

\maketitle

Bound states in the continuum (BICs)~\cite{Hsu2016,Koshelev2023} are remarkable quantum 
states whose hallmark is the absence of linewidth broadening and which therefore exhibit 
giant nonradiative lifetimes. Bound states embedded in the continuum have been first
suggested to exist by von Neumann and Wigner in the context of an open quantum system with 
rapidly oscillating potential~\cite{Wigner1929}. One century later, they became of 
practical relevance 
\cite{Marinica2008,Plotnik2011,Koshelev2020,Wu2024,Schiattarella2024,KoreshinPreprint2024} 
in photonics as particular solutions of the wave equation 
\cite{Monticone2014,Bulgakov2014,Sadrieva2019,Sadreev2021,Huang2023} and opened up
important applications \cite{Lei2023,Kang2023}.
However, the proposed quantum-mechanical examples of systems with BICs still fall
into the category of speculative theoretical work, rather than practical realization.
The theoretical setups \cite{Wigner1929,Stillinger1975,Fri85a,Fri85b} are very hard 
to implement in atomic systems as they require rather challenging conditions such as a 
complicated form of the potential, extremely strong external magnetic field, etc.
As a result, to our knowledge, so far there exists no quantum system that, under 
realistic conditions and at the current precision of measurements, allows one to identify 
BICs.

In this Letter, we propose a realistic semiconductor quantum system, which contains 
nontrivial BICs and which can be fabricated and studied experimentally in the near future. 
Trivial realizations of BICs can be easily obtained in semiconductor systems possessing a 
subband that differs in symmetry from lower subbands. In this case, the trivial BICs are 
uncoupled from the dominant decay channels of the lower subbands and thus have infinite 
lifetimes \cite{Landau,Scheuler2024}.
These symmetry-protected states have already been theoretically studied by investigating the
zero-linewidth states of electron-impurities in GaAs-based quantum 
wells (QWs) \cite{Schmelcher2005,Belov2022}.
By contrast, here we show the emergence of nontrivial BICs of electron-hole (eh) pairs 
confined in QWs. These BICs are coupled to other states, but exhibit zero 
linewidth broadening due to destructive interference of adjacent resonances 
as described by Friedrich and Wintgen \cite{Fri85b}.
We quantitatively show the appearance of the BICs and their 
nonzero-linewidth partner states by precise B-spline calculations of the complex-rotated 
Hamiltonian \cite{Moi98,Scheuler2024} for Rydberg excitons confined in cuprous oxide 
QWs \cite{Belov2024}.
We also provide a simple, but more approximate qualitative description
of BIC formation within the coupled-channel Schr\"{o}dinger equation
by means of quantum defect theory (QDT) \cite{Seaton1983,FriedrichBook}.

The theory is applied to describe BICs of eh pairs in cuprous oxide QWs.
Some of the already grown samples are hundreds of nanometers thick and
thus enable one to strongly confine only very large Rydberg
excitons~\cite{Naka2018,NakaPRL,Konzelmann2020}.
However, the thinnest fabricated Cu$_2$O films reported recently
\cite{Awal2024} are only 20-30~nm thick and are potentially suitable
for the study of lower exciton states.
In practice, the detection of exciton transitions in such systems is
quite complicated.
In addition to the radiative broadening, the exciton states are broadened
due to imperfections of the heterostructure, as well as interaction with 
phonons \cite{Fr1954,Schweiner16a}.
Nevertheless, large exciton oscillator strengths together with zero nonradiative broadening 
of BICs and the suppression of other parasitic processes opens up the possibility to 
experimentally detect these states.

It is worth noting that cuprous oxide QWs are not the only system that potentially
contains BICs. As shown previously \cite{Schmelcher2005}, a similar Hamiltonian describes 
GaAs-based heterostructures with QWs and other layered structures, too. However, 
the binding energy of excitons in bulk GaAs is about 4~meV and, thus, the corresponding 
avoided crossings of resonances are less than 0.1~meV wide \cite{Belov2019}. This
significantly complicates the experimental detection of BICs in GaAs-based heterostructures.
In contrast to GaAs, in cuprous oxide the large Rydberg energy of about 
90~meV \cite{Kaz14,Schweiner16b}, together with precise spectroscopic measurements allow 
one to firmly detect the splitting between the BIC and its partner state on the order of 
2~meV as predicted by our calculations. We expect that BICs can be experimentally detected 
in other layered semiconductor systems with confined excitons having large Rydberg 
energies \cite{He2014,Chernikov2014}. The BICs will facilitate the enhancement of optical 
nonlinearities and allow one to improve performance of exciton devices such as exciton 
lasers \cite{Kang2023}.
\begin{figure}
\includegraphics*[width=1.0\linewidth]{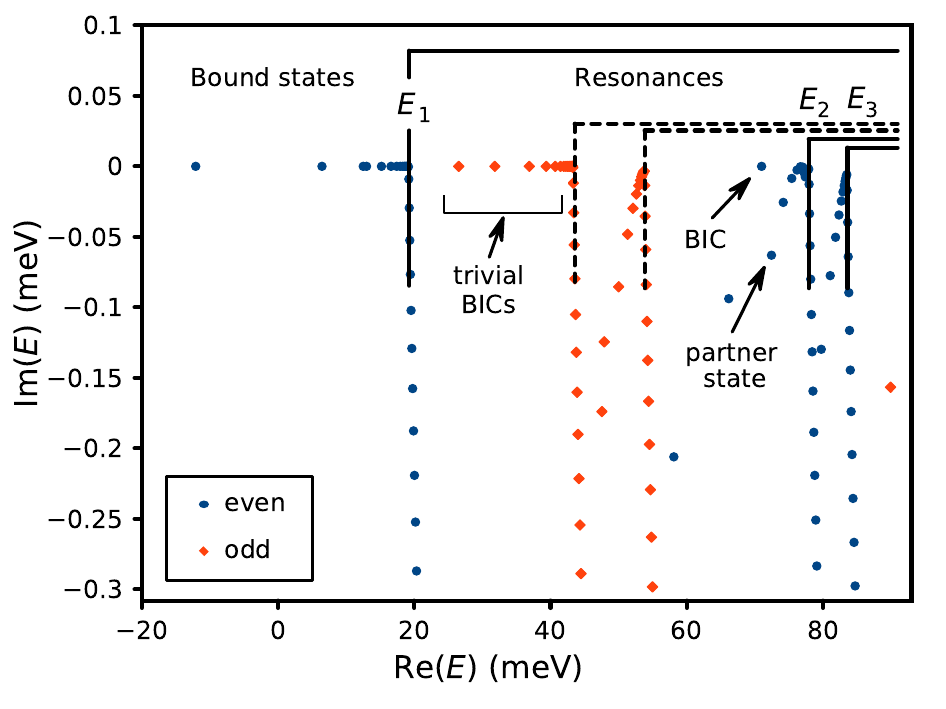}
\caption{Real and imaginary parts of the eh energies computed by the complex 
coordinate-rotation technique. The channels $E_{n}$ of even parity, combining e and h 
quantum-confinement subbands as well as the corresponding branches of the continuum, are 
enumerated by the index $n=1,2,3$. The bound states and resonances, including trivial 
(symmetry-protected) and nontrivial BICs, are explicitly denoted. The calculations were 
performed for a magnetic quantum number $m=1$ and a QW width $L=6.87$~nm for which 
a BIC at $E=70.98$~meV and its partner state appear.}
\label{figure1}
\end{figure}

For the theoretical description of eh pairs in a semiconductor QW we
use a hydrogenlike two-band model, as already introduced in
Refs.~\cite{Belov2019,Scheuler2024,Belov2024}.
The Hamiltonian
\begin{align}
\nonumber &H(z_{\mathrm{e}},z_{\mathrm{h}},\rho)=-\frac{\hbar^{2}}{2m_{\mathrm{e}}} 
\frac{\partial^{2}}{\partial z_{\mathrm{e}}^{2}}+V_{\mathrm{e}}(z_{\mathrm{e}})-
\frac{\hbar^{2}}{2m_{\mathrm{h}}} 
\frac{\partial^{2}}{\partial z_{\mathrm{h}}^{2}}+V_{\mathrm{h}}(z_{\mathrm{h}})\\
 &-\frac{\hbar^{2}}{2\mu}\left( \frac{\partial^{2}}{\partial \rho^{2}} + \frac{1}{\rho} 
 \frac{\partial}{\partial \rho} -\frac{m^{2}}{\rho^{2}} \right) -\frac{e^{2}}{\epsilon 
 \sqrt{\rho^{2}+(z_{\mathrm{e}}-z_{\mathrm{h}})^{2}}}
\label{eq:3Deq}
\end{align}
defines two subsystems: (i) an electron confined in a QW $V_{\mathrm{e}}
(z_{\mathrm{e}})$ and (ii) a hole confined in a QW $V_{\mathrm{h}}(z_{\mathrm{h}})$,
both coupled by the screened attractive Coulomb potential with
dielectric constant $\epsilon$ for both the QW and the barrier
material, i.e., the Hamiltonian~\eqref{eq:3Deq} does not consider a
dielectric contrast described by the Rytova-Keldysh (RK) potential
\cite{Rytova,Keldysh,Belov2024}.
The kinetic operators along the $z$ direction 
are defined by the effective masses $m_{\mathrm{e}}$ and $m_{\mathrm{h}}$ of electron (e) 
and hole (h), respectively. The cylindrical part of the energy operator is defined by the 
eh distance $\rho$ in the QW plane, and describes their relative motion 
($\mu=m_{\mathrm{e}}m_{\mathrm{h}}/(m_{\mathrm{e}}+m_{\mathrm{h}})$)
in this plane for a given angular harmonic specified by the magnetic quantum number $m$.
In addition to $m$, the parity $\pi_{z} = \pi_{z\mathrm{e}} \pi_{z\mathrm{h}} = \pm 1$,
related to the simultaneous exchange of $z_{\mathrm{e}} \to -z_{\mathrm{e}}$ and 
$z_{\mathrm{h}} \to -z_{\mathrm{h}}$, is also a good quantum number.

The resonances generated by the Hamiltonian~\eqref{eq:3Deq} can be identified by the 
complex coordinate-rotation technique \cite{Moi98}.
The rotation $\rho \to \rho\exp(i\theta)$ with an appropriate angle $\theta>0$ converts the 
outgoing scattering waves (as $\rho \to \infty$) to square-integrable functions allowing 
one to associate resonant states with the discrete spectrum of the non-Hermitian 
Hamiltonian $H\left(z_{\mathrm{e}},z_{\mathrm{h}},\rho\exp(i\theta) \right)$.
The artificially discretized continuum is then rotated into the complex lower half-plane by 
the angle $2\theta$.
The complex energies $E-i\Gamma/2$, that are independent of the angle 
of rotation, give energies and linewidths of the resonances, including
bound states for which $\Gamma\to 0$.
The rotation along only one variable $\rho$ is justified by the absence of scattering in 
the $z_{\mathrm{e}}$ and $z_{\mathrm{h}}$ directions due to the quantum-confinement 
barriers $V_{\mathrm{e}}(z_{\mathrm{e}})$ and $V_{\mathrm{h}}(z_{\mathrm{h}})$, which are 
taken to be large enough (or simply infinite) outside the QW of width $L$.
Details are given in Ref.~\cite{Scheuler2024}.
The eigenvalue problem for the complex-rotated Hamiltonian~\eqref{eq:3Deq} was numerically 
solved using an expansion of the wave function over a basis of high-order 
B-splines \cite{Bachau2001,Belov2024}.
\begin{figure}
 \includegraphics*[width=1.0\linewidth]{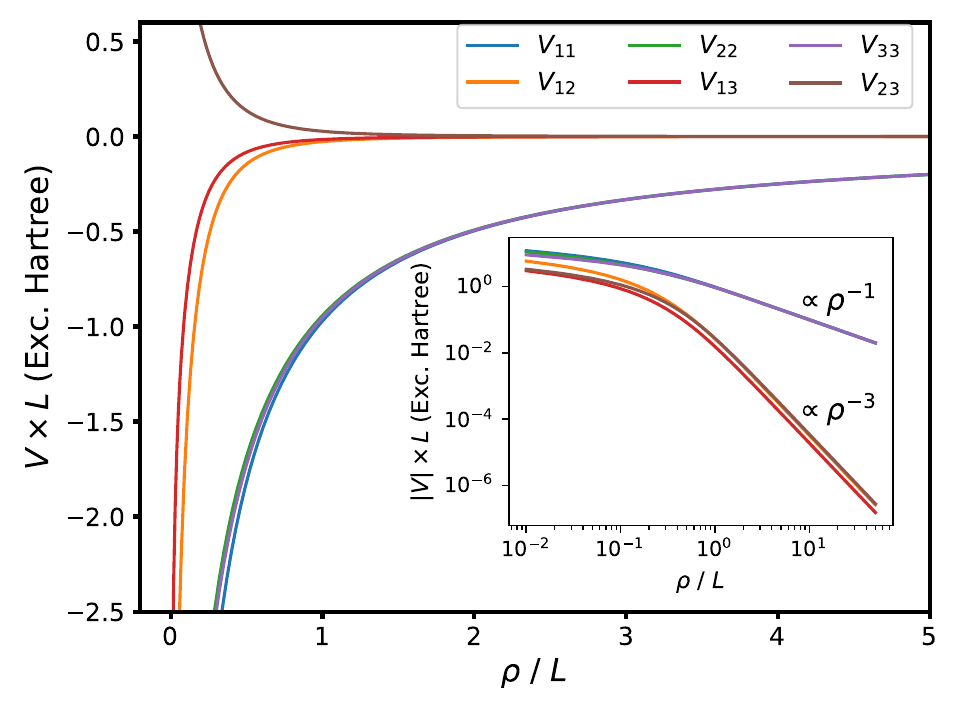}
\caption{\label{figure2}Asymptotic properties of the Coulomb coupling matrix element 
between different quantum-confinement states. The indices correspond to the even-parity 
channels combining the electron and hole quantum-confinement states. The inset shows the 
asymptotic behavior on a double logarithmic scale.
The curves for $V_{22}$ and $V_{33}$ visually overlap.}
\end{figure}

For a fixed QW width, the bound states, resonances, and rotated branches of the continua
are shown in the complex energy plane in Fig.~\ref{figure1}. The scattering thresholds of 
even parity (of e and h states) are denoted by $E_{n}$, $n=1,2,3$.
The energies of the excitons, i.e., the eh bound states, can be defined with respect to
the sum $E_{1}=E_{\mathrm{e}1}+E_{\mathrm{h}1}$ of the lowest energies of e and h in the 
confinement potentials. This sum specifies the lower boundary of the continuum, i.e., the 
lowest scattering threshold. Below $E_{1}$ there are only bound states of even parity, 
whereas above this threshold there are bound states of odd parity, the resonant states of 
even and odd parities. This is where potentially BICs can appear.
The bound states of odd parity can be considered as trivial, symmetry-protected BICs, see 
Fig.~\ref{figure1}. Above them, there are proper resonances with nonzero broadening, because
their wave functions have the same parity (symmetry) as the lower subband \cite{Belov2019} 
and, hence, they are coupled to the continuum.
Among the resonances, the BIC occurring at QW width $L=6.87\,$nm
and its partner state are explicitly denoted by arrows.

The resonances and BICs that appear for varying QW width $L$ can be studied in the 
framework of the coupled-channel Schr\"{o}dinger equation for the three lowest even-parity 
channels \cite{FriedrichBook}
\begin{align}
\nonumber & \left[ E^{}_{n} - \frac{\hbar^{2}}{2\mu}\left( \frac{\partial^{2}}{\partial 
\rho^{2}} + \frac{1}{\rho} \frac{\partial}{\partial \rho} -\frac{m^{2}}{\rho^{2}} \right) + 
V_{nn}(\rho) \right] \Psi_{n}(\rho) \\
& + \sum_{n' \neq n} V_{nn'}(\rho) \Psi_{n'}(\rho) = E \Psi_{n}(\rho).
\label{eq:CC}
\end{align}
The terms in square brackets in Eq.~\eqref{eq:CC} represent the diagonal elements of the 
matrix equation. For the potentials 
\begin{align}
&V_{nn'}(\rho) = V_{\{ij\},\{kl\}}(\rho)\nonumber \\
&=-\frac{e^{2}}{\epsilon}\iint\frac{\psi_{\mathrm{e}i}(z_{\mathrm{e}}) \psi_{\mathrm{h}j}
(z_{\mathrm{h}}) \psi_{\mathrm{e}k}(z_{\mathrm{e}}) \psi_{\mathrm{h}l}(z_{\mathrm{h}})}
{\sqrt{\rho^{2}+(z_{\mathrm{e}}-z_{\mathrm{h}})^{2}}} \, dz_{\mathrm{e}} \, dz_{\mathrm{h}}
\label{eq:VC}
\end{align}
we use numerically exact expansions, which is a pre\-re\-quisite for the rather accurate 
QDT \cite{Seaton1983} results presented below. More details are given in the Supplemental
Material~\cite{sm}.
The deviation from the exact Rydberg series originates from the non-diagonal Coulomb 
coupling of quantum-confinement states that are combined into channels of the same (even) 
parity indexed over energy from the bottom to the top by $n=\{i,j\}$, where $i$ and $j$ 
enumerate e and h single-particle states.
The non-diagonal Coulomb coupling also introduces the avoided crossings of the
bound states and resonances of the same parity as functions of $L$.

\begin{figure}
\includegraphics*[width=1.0\columnwidth]{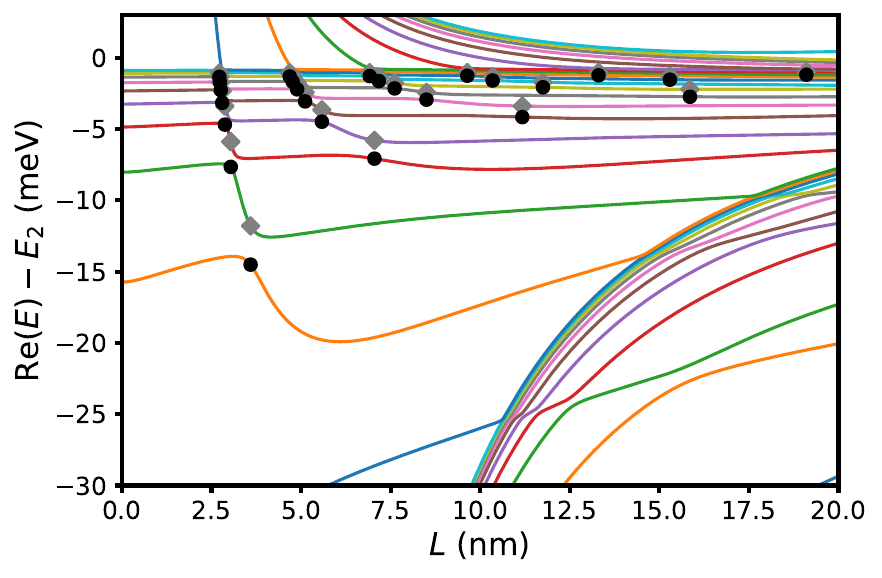}
\caption{\label{figure3}Energies of the even-parity resonances with respect to the 
threshold $E^{}_{2}$ as functions of the QW width $L$. The black circles and grey diamonds 
denote the BICs and their partner states, respectively.}
\end{figure}
As shown in Fig.~\ref{figure2}, the diagonal potential terms $V_{nn}(\rho)$ scale as 
$\rho^{-1}$, whereas the non-diagonal $V_{nn'}(\rho)$, with $n\neq n'$, decrease as 
$\rho^{-3}$. Therefore, asymptotically (as $\rho \to \infty$) the channels decouple. 
This allows one to obtain the scattering wave functions $\Phi_1(\rho)$
in the open channel ($n=1$) from Eq.~\eqref{eq:CC} using only the diagonal potential
$V_{11}(\rho)$.
We use $\Phi_n(\rho)$ for the notation of the one-channel wave functions to distinguish
them from the solutions $\Psi_n(\rho)$ of the coupled-channel Schrödinger 
equation~\eqref{eq:CC}, i.e., including the non-diagonal potentials.

For the two closed channels ($n=2,3$), the one-channel QDT wave functions $\Phi_n^N(\rho)$
with principal quantum number $N$ are obtained as solution of the diagonal part of
Eq.~\eqref{eq:CC}. They form a basis to set up a matrix representation of the 
coupled-channel Schrödinger equation, yielding the multi-channel wave functions.
Details are given in the Supplemental Material~\cite{sm}.
With the solution of Eq.~\eqref{eq:CC} at hand, the linewidth broadenings can 
be estimated by the Feshbach formula \cite{Feshbach1958,Schmelcher2005}
\begin{equation}
\Gamma = 2\pi \left| \sum_{n=2}^3 \langle \Phi_{1} | V_{1n} | \Psi_{n} \rangle \right|^{2},
\label{eq:Feshbach}
\end{equation}
which describes the overlap of the coupling potential between the components $\Psi_{n}$, 
associated with the second and the third channels of the complete solution, 
and the regular scattering solution $\Phi_{1}$ in only the first channel.
It represents the interaction of the two resonances with the continuum background.
The BIC appears when $\Gamma$ vanishes due to destructive interference of the
matrix elements in Eq.~\eqref{eq:Feshbach}.

%
\begin{figure}
\includegraphics*[width=1.0\columnwidth]{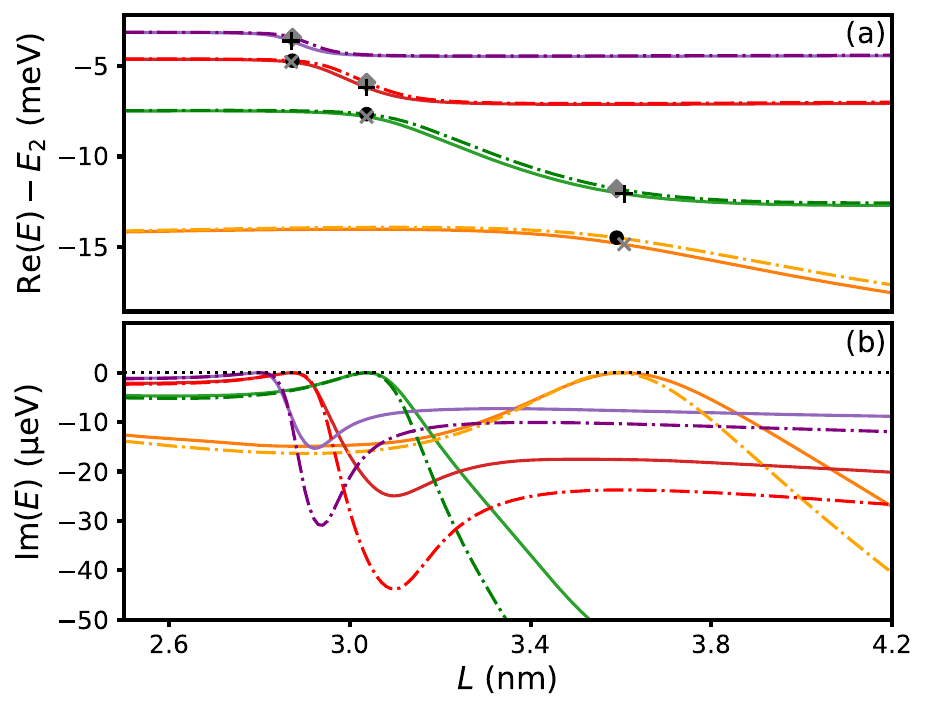}
\caption{\label{figure4}Real and imaginary parts of energies of the resonances with 
respect to the threshold $E_{2}$ as functions of QW width $L$. The linewidth is 
$\Gamma=-2 \Im(E)$.
The color code is the same as in Fig.~\ref{figure3}. The solid curves show the results of 
complex coordinate-rotation, whereas the dashed curves are the solutions of the 
coupled-channel Schr\"{o}dinger equation using QDT. The zero linewidths indicate the 
emergence of BICs.}
\end{figure}
%
Resonant energies of the Rydberg series as a function of the QW width $L$ are shown in 
Fig.~\ref{figure3}. They are plotted with respect to the $E^{}_{2}$ threshold.
The energies of this subband tend to lie horizontally along the threshold, whereas upper 
(lower) subband energies are decreasing (increasing) as $L \to \infty$.
Due to their same even parity, this subband is coupled to the continuum of subband $E_{1}$ 
and thus two resonances of similar energy exhibit avoided crossings as $L$ varies.
The calculated data are additionally given in Table~\ref{tab:results_table}.
Figure~\ref{figure4} shows in detail the dependencies on $L$ of real and imaginary parts of 
several resonance energies.
\begin{table}
    \centering
    \caption{Energies of the BICs and the partner states obtained from QDT and B-spline 
    calculations. The values are given with respect to the $E_{2}$ threshold.}
\begin{tabular}{rrr|rrr}
\hline
\multicolumn{3}{c|}{QDT} & \multicolumn{3}{c}{B-splines} \\
\hline
&BIC&\multicolumn{1}{r|}{partner}&&BIC&\multicolumn{1}{r}{partner} \\
\hline
     $L$ (nm) & $E$ (meV) & $E$ (meV) & $L$ (nm) & $E$ (meV) & $E$ (meV) \\
\hline 
$3.590$ & $-14.505$ & $-11.793$ & $3.607$ & $-14.875$ & $-12.075$ \\
$3.037$ & $-7.652$  & $-5.890$  & $3.037$ & $-7.793$  & $-6.174$ \\
$7.047$ & $-7.072$  & $-5.794$  & $6.871$ & $-7.362$  & $-5.911$ \\
$2.873$ & $-4.686$  & $-3.392$  & $2.870$ & $-4.754$  & $-3.593$ \\
$5.580$ & $-4.484$  & $-3.651$  & $5.534$ & $-4.614$  & $-3.763$ \\
$11.174$ & $-4.161$ & $-3.389$  & $11.293$ & $-4.418$ & $-3.557$ \\
$2.799$ & $-3.158$  & $-2.317$  & $2.796$ & $-3.199$  & $-2.399$ \\
$5.112$ & $-3.059$  & $-2.424$  & $5.077$ & $-3.132$  & $-2.530$ \\
$8.499$ & $-2.937$  & $-2.430$  & $8.564$ & $-2.673$  & $-2.498$ \\
$15.855$ & $-2.736$ & $-2.210$  & $15.204$ & $-2.549$ & $-2.267$ \\
$2.759$ & $-2.271$  & $-1.725$  & $2.760$ & $-2.297$  & $-1.761$ \\
$4.889$ & $-2.213$  & $-1.738$  & $4.858$ & $-2.260$  & $-1.814$ \\
$7.610$ & $-2.150$  & $-1.765$  & $7.307$ & $-2.217$  & $-1.766$ \\
$11.750$ & $-2.071$ & $-1.718$  & $11.407$ & $-2.179$ & $-1.803$ \\
$2.707$ & $-1.070$  & $-0.878$  & $2.731$ & $-1.729$  & $-1.360$ \\
$4.761$ & $-1.674$  & $-1.331$  & $4.733$ & $-1.706$  & $-1.377$ \\
$7.172$ & $-1.636$  & $-1.337$  & $6.979$ & $-1.680$  & $-1.363$ \\
\hline
\end{tabular}
\label{tab:results_table}
\end{table}
One observes a good agreement of the energies as well as positions of the BICs calculated 
by both methods: the solid curves represent the numerically exact B-spline calculations and 
the dashed curves are the approximate results by three-channel QDT.
The agreement of the absolute values of the linewidth broadenings is more qualitative due 
to the limited number of channels considered for the QDT solution of Eq.~\eqref{eq:CC}.
Nonetheless, both precise and approximate calculations show that for some values of $L$ the 
imaginary parts of energies (i.e., the linewidths) vanish and BICs appear.
The linewidths of the partner states also change with varying $L$, but they remain nonzero 
in the vicinity of the considered avoided crossing.
Interestingly, the local maxima of linewidth broadenings of the partner states are realised 
not exactly at the location of the BIC, but they are slightly shifted. This fact was also 
noted in Ref.~\cite{Schiattarella2024} and originates from their interaction with other 
resonances and subbands, complicating the problem more than when it was first discussed by 
Friedrich and Wintgen \cite{Fri85b}.

The idea of Friedrich and Wintgen had been that the BIC appears due to the destructive 
interference of two coupled resonances both in the continuum of a much lower open channel
(subband). Here we give an alternative and more intuitive derivation of this effect.
When restricting the QDT basis to only one state $\Phi_n^N(\rho)$ in each closed channel,
those energies undergo an avoided crossing when the QW width $L$ is varied,
all calculations can be done analytically.
With $E_\mathrm{I}$ and $E_\mathrm{II}$ the energies of the two
selected states $\Phi_2^\mathrm{I}$ and $\Phi_3^\mathrm{II}$ in the two 
closed channels ($n=2,3$), and $W=\langle\Phi_2^\mathrm{I}|V_{23}|\Phi_3^\mathrm{II}\rangle$
the matrix element of the coupling potential, the resulting $(2\times 2)$ Hamiltonian 
matrix can be diagonalized by an orthogonal transformation, i.e.
\begin{equation}
O^T
  \left( {\begin{array}{cc}
   E_\mathrm{I} & W\\
   W & E_\mathrm{II}\\
  \end{array} } \right)
O
= \left( {\begin{array}{cc}
   E_{-} & 0\\
   0 & E_{+}\\
  \end{array} } \right)
\end{equation}
with
\begin{equation}
O =
  \left( {\begin{array}{rr}
   \cos\alpha & \sin\alpha\\
   -\sin\alpha & \cos\alpha\\
  \end{array} } \right) .
\label{eq:orth_trans}
\end{equation}
The angle $\alpha$, given by
\begin{equation}
  \tan\alpha = \frac{1}{2W}\left[\sqrt{(E_\mathrm{II}-E_\mathrm{I})^2+4W^2}-(E_\mathrm{II}-E_\mathrm{I})\right] \, ,
\label{eq:tan_alpha}
\end{equation}
defines the mixing of states and specifies the energy eigenvalues
\begin{equation}
  E_\pm = \frac{1}{2}\left[(E_\mathrm{I}+E_\mathrm{II})\pm\sqrt{(E_\mathrm{II}-E_\mathrm{I})^2+4W^2}\right] \, .
\end{equation}
Using the eigenvectors given by the columns of the matrix~\eqref{eq:orth_trans},
the Feshbach formula~\eqref{eq:Feshbach} yields
\begin{subequations}
\label{eq:Gamma}
\begin{align}
  \Gamma_{-} &= \left(\sqrt{\Gamma_\mathrm{I}}\cos\alpha-\sqrt{\Gamma_\mathrm{II}}\sin\alpha\right)^2\, ,
\label{eq:Gamma_-}\\
  \Gamma_{+} &= \left(\sqrt{\Gamma_\mathrm{I}}\sin\alpha+\sqrt{\Gamma_\mathrm{II}}\cos\alpha\right)^2\, ,
\label{eq:Gamma_+}
\end{align}
\end{subequations}
with $\Gamma_\mathrm{I}=2\pi|\langle \Phi_{1}| V_{12}|\Phi_2^\mathrm{I} \rangle|^{2}$
and $\Gamma_\mathrm{II}=2\pi|\langle \Phi_{1}| V_{13}|\Phi_3^\mathrm{II}\rangle|^{2}$
the linewidths of the one-channel states $\Phi_2^\mathrm{I}$ and $\Phi_3^\mathrm{II}$.
The condition for complete destructive interference of the linewidths becomes obvious 
from Eq.~\eqref{eq:Gamma_-}: a BIC with $\Gamma_{-}=0$ occurs at energy $E=E_{-}$ when
$\tan\alpha=\sqrt{\Gamma_\mathrm{I}/\Gamma_\mathrm{II}}$.
This yields the condition \cite{Fri85b}
\begin{equation}
    (\Gamma_\mathrm{II}-\Gamma_\mathrm{I})W = \sqrt{\Gamma_\mathrm{I}\Gamma_\mathrm{II}}(E_\mathrm{II}-E_\mathrm{I}) \; .
\end{equation}
From Eqs.~\eqref{eq:Gamma} it follows that 
$\Gamma_{-}+\Gamma_{+}=\Gamma_\mathrm{I}+\Gamma_\mathrm{II}$.
Thus, the partner state of a BIC with $\Gamma_{-}=0$ has the linewidth 
$\Gamma_{+}=\Gamma_\mathrm{I}+\Gamma_\mathrm{II}$, i.e., the sum of the linewidths of the 
two involved one-channel states.

As discussed above, the QW widths and energies of BICs in Fig.~\ref{figure4}
and Table~\ref{tab:results_table} obtained from QDT and the B-spline
computations are in remarkably good agreement.
Nevertheless, our model shows only the principal effect which may be
shadowed in a real experiment due to simplifications in the Hamiltonian~\eqref{eq:3Deq}.
On the one hand, the parabolic valence band of cuprous oxide
is just a (first order) approximation, representing the dominant
diagonal part of the Suzuki-Hensel Hamiltonian \cite{Suzuki1974,Schweiner17b},
allowing us to use the hydrogenlike two-band model in Eq.~\eqref{eq:3Deq}.
For bulk cuprous oxide the impact of the complete Suzuki-Hensel valence band structure
on the yellow exciton series has been investigated both experimentally
and theoretically \cite{Kaz14,Thewes15,Schoene16,Schweiner16b},
however, to our knowledge not yet for cuprous oxide in quantum wells.
For relatively narrow QWs and for highly excited Rydberg states, the
quantum confinement dominates and the effects of the complex band
structure can be considered as a small perturbation~\cite{Belov2024}.
On the other hand, in the Hamiltonian~\eqref{eq:3Deq} we have assumed
the same dielectric constant $\epsilon=7.5$ for the QW and the barrier
material, whereas different values $\epsilon_{\mathrm{QW}} \ne \epsilon_{\mathrm{b}}$
are more realistic for experiments with thin films.
The effect of the dielectric contrast has been investigated by
Rytova and Keldysh \cite{Rytova,Keldysh}.
In Ref.~\cite{Belov2024} it has been shown that the dielectric
contrast with $\epsilon_{\mathrm{b}} < \epsilon_{\mathrm{QW}}$
described by the RK potential causes a downshift of
energies, however, the structure of energy levels is similar for both
the Coulomb and the RK potential.
The symmetry introduced by the quantum confinement is kept
independent of the non-parabolicity of the band structure
and the different dielectric constants.
As a result, the avoided crossings between exciton resonances belonging to different channels,
and thus the occurrence of BICs via destructive interference of these
states persists.
We therefore expect that BICs will be observed in future experiments
on thin cuprous oxide films at quantum well widths and energies comparable with our predictions.

In conclusion, we emphasize that, although BICs have been experimentally realized in 
photonics, atomic quantum systems seem to be inappropriate for their detection.
Instead, we propose a Rydberg-exciton semiconductor quantum system which can, in 
principle, be fabricated down to lattice constant precision \cite{Naka2018,NakaPRL}. 
Due to their giant exciton oscillator strengths \cite{Kaz14} and large enough splitting 
of BICs and their partner states, these state pairs can be spectroscopically resolved. 
Here, we have rigorously shown the principal phenomenon of BIC emergence, however,
more research is required to take into account all the features of the 
system \cite{Rytova,Keldysh,Schweiner16b,Belov2024}. We believe that the experimental 
detection of BICs in cuprous oxide will open up ways to enhance optical nonlinearities 
and to improve the performance of exciton lasers \cite{Kang2023}.

\acknowledgments
This work was supported by Deutsche Forschungsgemeinschaft (DFG) through the DFG Priority 
Programme 1929 ``Giant interactions in Rydberg Systems'' (GiRyd), Grant Nos.\ MA 1639/16-1 
and SCHE 612/4-2.

%

\end{document}


\title{Supplemental Material for\\ ``Bound states in the continuum in cuprous oxide quantum wells''}
\author{Angelos Aslanidis}
\author{Jörg Main}
\email[Email: ]{main@itp1.uni-stuttgart.de}
\author{Patric Rommel}
\affiliation{Institut für Theoretische Physik I, Universität Stuttgart, 70550 Stuttgart, Germany}
\author{Stefan Scheel}
\author{Pavel A. Belov}
\affiliation{Institut für Physik, Universität Rostock,
  Albert-Einstein-Straße 23-24, 18059 Rostock, Germany}


\maketitle

In this Supplemental Material, we present the derivation of the
potentials $V_{nn'}(\rho)$ in Eq.~(3) of the main text and explain
the application of quantum defect theory (QDT) to excitons in cuprous
oxide quantum wells.

\section{Derivation of the diagonal and coupling potentials}
A precise computation of the potentials $V_{nn'}(\rho)$ in the
coupled-channel Schrödinger equation~(2) of the main text is the
prerequisite for obtaining accurate results for the bound states in
the continuum (BICs) by application of quantum defect theory.
The potentials (in exciton-Hartree units with $e=4\pi\epsilon_0\epsilon=1$)
are given as
\begin{align}
&V_{nn'}(\rho) = V_{\{ij\},\{kl\}}(\rho)
  =-\int_{-L/2}^{L/2}\int_{-L/2}^{L/2}
    \frac{\psi_{\mathrm{e}i}(z_{\mathrm{e}}) \psi_{\mathrm{h}j}(z_{\mathrm{h}}) \psi_{\mathrm{e}k}(z_{\mathrm{e}})
    \psi_{\mathrm{h}l}(z_{\mathrm{h}})}{\sqrt{\rho^{2}+(z_{\mathrm{e}}-z_{\mathrm{h}})^{2}}}
    \, dz_{\mathrm{e}} \, dz_{\mathrm{h}} \, ,
\label{eq:V_nm}
\end{align}
where $\psi_{\mathrm{e}i}(z_{\mathrm{e}})$, $\psi_{\mathrm{h}j}(z_{\mathrm{h}})$, 
$\psi_{\mathrm{e}k}(z_{\mathrm{e}})$, and
$\psi_{\mathrm{h}l}(z_{\mathrm{h}})$ are the (analytically known) 
states of electron
and hole in the quantum well with width $L$.
As an example, we derive the diagonal potential $V_{11}(\rho)$, where both 
the electron and the hole are in the quantum well ground state, i.e.
\begin{equation}
  \psi_{\mathrm{e}1}(z_{\mathrm{e}})=\sqrt{\frac{2}{L}}\cos\left(\frac{\pi z_{\mathrm{e}}}{L}\right) \, , \quad
  \psi_{\mathrm{h}1}(z_{\mathrm{h}})=\sqrt{\frac{2}{L}}\cos\left(\frac{\pi z_{\mathrm{h}}}{L}\right) \, .
\label{eq:psi1}
\end{equation}
Using the scaling $\tilde\rho=\pi\rho/L$, $\tilde z_{\mathrm{e,h}}=\pi z_{\mathrm{e,h}}/L$,
and center and relative coordinates $Z=(\tilde z_{\mathrm{e}}+\tilde z_{\mathrm{h}})/2$,
$z=\tilde z_{\mathrm{e}}-\tilde z_{\mathrm{h}}$, we obtain
\begin{align}
  V_{11}(\rho) &= -\frac{4}{\pi L}\int_{-\pi}^{\pi}dz \int_{-(\pi-|z|)/2}^{(\pi-|z|)/2}dZ
  \frac{1}{\sqrt{\tilde\rho^2+z^2}}\cos^2(Z+z/2)\cos^2(Z-z/2)\nonumber\\
  &= -\frac{4}{\pi L}\int_0^\pi dz \frac{1}{\sqrt{\tilde\rho^2+z^2}}
    \frac{1}{8}[3\sin(2z)+2(\pi-z)(\cos(2z)+2)]\nonumber\\
  &\stackrel{z\equiv\tilde\rho x}{=} -\frac{1}{2\pi L}\int_0^{\pi/\tilde\rho}dx
    \frac{1}{\sqrt{1+x^2}}[3\sin(2\tilde\rho x)+2(\pi-\tilde\rho x)(\cos(2\tilde\rho x)+2)] \, .
\label{eq:V11}
\end{align}
The remaining integral cannot be solved analytically.
We now consider two cases, viz.\ the regions $\rho\gtrsim 1.5L$
and $\rho\lesssim 1.5L$.
For $\rho\gtrsim 1.5L$, i.e., large values $\tilde\rho\gtrsim 1.5\pi$,
the values of $x$ in Eq.~\eqref{eq:V11} are restricted to $|x|<1$, and
thus the Taylor series
\begin{align}
  \frac{1}{\sqrt{1+x^2}} = 1 - \frac{1}{2}x^2 + \frac{3}{8}x^4 -
  \frac{5}{16}x^6 + \frac{35}{128}x^8 \mp\dots
\end{align}
converges.
Inserting this series into Eq.~\eqref{eq:V11} the integrals of all
individual terms can now be solved analytically providing the
asymptotic formula valid for $\rho\gtrsim 1.5L$
\begin{align}
  V_{11}(\rho) &= -\sum_{n=0}^{n_{\max}} c_n\frac{L^{2n}}{\rho^{2n+1}}
\label{eq:V11_expand1}
\end{align}
with in principle analytically known coefficients, see Table~\ref{tab:coefficients}.
\begin{table}
    \centering
    \caption{Coefficients of the asymptotic expansion~\eqref{eq:V11_expand1} of
    the diagonal potential $V_{11}(\rho)$.}
\begin{tabular}{ll}
\hline 
 $c_{0}$ & $\phantom{+}1$ \\
 $c_{1}$ & $-0.03267274151216445$ \\
 $c_{2}$ & $\phantom{+}0.004328401188903151$ \\
 $c_{3}$ & $-0.0009619175433854251$ \\
 $c_{4}$ & $\phantom{+}0.0002864777243269292$ \\
 $c_{5}$ & $-0.0001035053259902042$ \\
 $c_{6}$ & $\phantom{+}0.0000429546345632382$ \\
 $c_{7}$ & $-0.00001979479166677419$ \\
 $c_{8}$ & $\phantom{+}9.903495293081103\times 10^{-6}$ \\
 $c_{9}$ & $-5.294507909335873\times 10^{-6}$ \\
 $c_{10}$ & $\phantom{+}2.98956638899579\times 10^{-6}$ \\
\hline
\end{tabular}
\label{tab:coefficients}
\end{table}
We used $n_{\max}=10$ in our computations.
For $\tilde\rho\lesssim 1.5\pi$ the evaluation of
Eq.~\eqref{eq:V11} is less straightforward.
When writing
\begin{equation}
  V_{11}(\rho) = -\frac{1}{2\pi L}\int_0^{\pi/\tilde\rho}dx \frac{f(x)}{\sqrt{1+x^2}}
\label{eq:V11_f}
\end{equation}
the function
$f(x)=3\sin(2\tilde\rho x)+2(\pi-\tilde\rho x)(\cos(2\tilde\rho x)+2)$
can be expanded in terms of Legendre polynomials, transformed from the
standard interval $[-1,1]$ to the range $[0,\pi/\tilde\rho]$.
For $f(x)$ given as a polynomial, the integral in Eq.~\eqref{eq:V11_f} 
can be solved analytically.
Using Legendre polynomials up to order six yields the expansion
\begin{align}
  V_{11}(\rho) = &-\frac{1}{L}\bigg[d_1\frac{\rho}{L} + d_3\frac{\rho^3}{L^3} + d_5\frac{\rho^5}{L^5}
       + \left(d_0 + d_2\frac{\rho^2}{L^2} + d_4\frac{\rho^4}{L^4}\right)\sqrt{1+\rho^2/L^2}\nonumber\\
      &+ \left(e_0 + e_2\frac{\rho^2}{L^2} + e_4\frac{\rho^4}{L^4} + e_6\frac{\rho^6}{L^6}\right)
         \arcsinh\left(\frac{L}{\rho}\right)\bigg]
\label{eq:V11_expand2}
\end{align}
with coefficients $d_i$ and $e_i$.
This expansion is valid for $\rho\lesssim 1.5L$.
Note that contrary to the asymptotic formula~\eqref{eq:V11_expand1}
the coefficients in Eq.~\eqref{eq:V11_expand2} are not unique but
depend on the order of the expansion.
We used symbolic calculations with \textsf{Wolfram Mathematica}~\cite{Mathematica} for the
evaluation of Eq.~\eqref{eq:V11} and the computation of the
coefficients in Eqs.~\eqref{eq:V11_expand1} and \eqref{eq:V11_expand2}.

When the ground state wave functions~\eqref{eq:psi1} for the electron
and hole in the quantum well are replaced in Eq.~\eqref{eq:V_nm} with
appropriate excited states $\psi_{\mathrm{e}i,\mathrm{h}j}(z_{\mathrm{e,h}})$, similar expansions
as Eqs.~\eqref{eq:V11_expand1} and \eqref{eq:V11_expand2} for
$V_{11}(\rho)$ can be derived for other coupling potentials
$V_{nn'}(\rho)=V_{\{ij\},\{kl\}}(\rho)$ with $n>1$, $n'>1$, which are shown
in Fig.~2 of the main text.
The leading coefficient in Eq.~\eqref{eq:V11_expand1} is $c_0=1$ in case
of the diagonal potentials and $c_0=0$ in case of the non-diagonal coupling potentials.
This leads to the asymptotic behavior of the potentials for large
values of $\rho$ as discussed in the main text.

\section{Application of Quantum Defect Theory}
When applying QDT to cuprous oxide excitons in quantum wells, we only
consider the three lowest even-parity channels formed by quantum-confinement
quantum numbers $(n_{\mathrm{e}},n_{\mathrm{h}})=(1,1)$ for the first channel ($n=1$), and
$(n_{\mathrm{e}},n_{\mathrm{h}})=(3,1)$ and $(2,2)$ for the second ($n=2$) and third ($n=3$)
channels, respectively.
The restriction to only three channels is justified by the fact that
the threshold of the fourth even-parity channel with quantum
numbers $(n_{\mathrm{e}},n_{\mathrm{h}})=(1,3)$ is energetically well separated from the
three lower thresholds for quantum well widths $L\lesssim 20\,$nm.
In a first step we solve the Schrödinger equation~(2) in the main
text by considering only the diagonal potentials $V_{nn}(\rho)$.
In a second step, the lowest 10 eigenstates in each channel are then
taken to set up an eigenvalue problem by including the
nondiagonal coupling potentials $V_{nn'}(\rho)$.
The linewidths of the states obtained by the three-channel QDT 
are computed by the Feshbach formula, see below.
Finally, we search for states with vanishing linewidths, viz.\
the bound states in the continuum (BICs).

\subsection{One-channel QDT}
\begin{figure}
    \centering
    \includegraphics[width=0.6\linewidth]{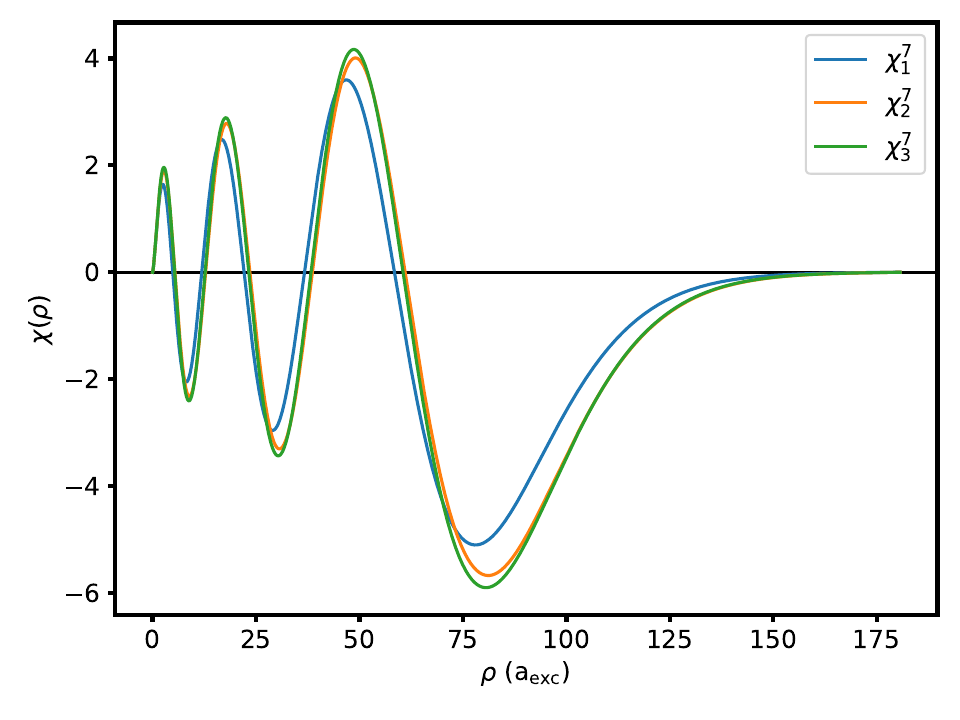}
    \caption{One-channel QDT wave functions 
    $\chi_n^N(\rho)=\sqrt{\rho}\Phi_n^N(\rho)$
    with angular quantum number $m=1$ and principal quantum number $N=7$
    in the even-parity channels $n=1$ to $3$.
    $a_\mathrm{exc}$ is the exciton Bohr radius.}
    \label{fig:sm_figure1}
\end{figure}

The wave functions $\Phi_n^N(\rho)$ with principal quantum number $N$
in the even-parity channel $n$ described by the diagonal potential
$V_{nn}(\rho)$ and threshold energy $E_n$ are solutions of the one-channel
quantum-defect Schrödinger equation
\begin{equation}
  \left[ E_n - \frac{\hbar^{2}}{2\mu}\left( \frac{\partial^{2}}{\partial 
  \rho^{2}} + \frac{1}{\rho} \frac{\partial}{\partial \rho} -\frac{m^{2}}{\rho^{2}}
  \right) + V_{nn}(\rho) \right]\Phi_n^N(\rho) = E_n^N \Phi_n^N(\rho) \, .
  \label{eq:H_1channel}
\end{equation}
They are obtained by numerical integration
of the wave function 
with initial conditions
\begin{align}
  \Phi_n^N(\rho_0) = \rho_0^{|m|} \; ,\quad
  \frac{d}{d\rho}\Phi_n^N(\rho_0) = |m|\, \rho_0^{|m|-1}
  \label{eq:init_cond}
\end{align}
at a small value of $\rho_0$ to fulfill the boundary condition
$\Phi_n^N(\rho)\sim\rho^{|m|}$ in the limit $\rho\to 0$.
Here, $m$ is the magnetic quantum number.
To also fulfill the boundary condition $\Phi_n^N(\rho)=0$ for
$\rho\to\infty$, a numerical root search is applied by varying the
energy $E=-E_{\mathrm{Ryd}}/N_{\mathrm{eff}}^2$, with the effective
quantum numbers $N_{\mathrm{eff}}=N-\delta_N$ and the quantum defects
$\delta_N$.
For illustration, the one-channel QDT wave functions
$\chi_n^N(\rho)=\sqrt{\rho}\Phi_n^N(\rho)$ with angular quantum 
number $m=1$ and principal quantum number $N=7$ in the even-parity 
channels $n=1$ to $3$ are shown in Fig.~\ref{fig:sm_figure1}.

\subsection{Three-channel QDT}
\label{sec:three-channel-QDT}
The wave functions $\Phi_n^N(\rho)$ with the magnetic quantum number $m=1$
and the principal quantum numbers $N=2$ to $11$ in the three lowest
even-parity channels ($n=1,2,3$) are taken as a basis for the
expansion
\begin{equation}
    \Psi_{n}(\rho)= \sum_{N} c_{n}^{N} \Phi_{n}^{N}(\rho)
\label{eq:psi_expand}
\end{equation}
of the solution of the coupled-channel Schrödinger equation~(2)
in the main text.
The nondiagonal matrix elements of the $(30\times 30)$-dimensional
Hamiltonian matrix are obtained by numerical integration using the
coupling potentials $V_{nn'}(\rho)$ with $n\ne n'$ and the wave
functions $\Phi_n^N(\rho)$ of the three lowest channels.
The Hamiltonian matrix is diagonalized with a conventional linear
algebra routine to obtain the energy eigenvalues presented in Figs.~3
and 4 of the main text.
The coefficients $c_n^N$ of the eigenvectors of the $(30\times 30)$-dimensional
Hamiltonian matrix can be used to expand the final wave functions
$\Psi$ as solution of the coupled-channel Schrödinger equation~(2) in
the main text in terms of the one-channel QDT basis functions
$\Phi_n^N(\rho)$.

\subsection{Computation of the linewidths}
The linewidths of resonances in the energy region with one open
channel ($E_1 < E < E_2$) are given by the
Feshbach formula~\cite{Feshbach1958,FriedrichBook}
\begin{equation}
\Gamma = 2\pi |\langle \Phi_{1} |V_{12}\Psi_{2}+V_{13}\Psi_{3}\rangle|^2 \, ,
\label{eq:Gamma}
\end{equation}
where $(\Psi_{2},\Psi_{3})^{T}$ is a part of the solution of the three-channel Schrödinger 
equation associated with the two closed channels.
The energy normalized regular scattering wave function $\Phi_{1}(\rho)$ in the open channel 
($n=1$) is obtained by numerical integration of the wave function with the initial 
conditions~\eqref{eq:init_cond} and is adjusted to the asymptotic form~\cite{Landau}
\begin{equation}
    \Phi_{1}^{\mathrm{asy}}(\rho) = \frac{1}{\sqrt{\hbar \pi}} \left(\frac{2m}{E} \right)^{1/4} \frac{1}{\sqrt{\rho}}\sin{\left(k\rho+\frac{1}{k}\ln{(2k\rho)}-\pi/2+\delta(k)\right)},
\label{eq:phi_asy}
\end{equation}
where $\delta(k)$ is a phase shift and $k=\sqrt{2mE/\hbar^{2}}$,
to guarantee proper normalization.
Using Eq.~\eqref{eq:psi_expand} for the expansion of the wave function
$\Psi$ in terms of the one-channel QDT basis functions $\Phi_n^N(\rho)$,
the overlap matrix elements in Eq.~\eqref{eq:Gamma} are obtained as 
\begin{equation}
  \langle \Phi_{1} |V_{12}\Psi_{2}+V_{13}\Psi_{3}\rangle =
  \sum_{N} \left[
    c_2^N \langle \Phi_{1}|V_{12}|\Phi_2^N\rangle +
    c_3^N \langle \Phi_{1}|V_{13}|\Phi_3^N\rangle \right] \, .
\label{eq:overlap}
\end{equation}
A vanishing overlap matrix element~\eqref{eq:overlap} at a certain
width $L$ of the quantum well, due to destructive interference of
the wave functions in the two closed channels, characterizes a BIC.
When the basis of one-channel QDT wave functions for the
matrix setup is restricted to only
one state in each of the two closed channels, which undergo an avoided
crossing when the QW width $L$ is varied, the energy eigenvalues,
linewidths, and the condition for the occurrence of a BIC are obtained
analytically as described in the main text.

%